%% file: main.tex
\documentclass[sigconf]{acmart}
\acmConference[SE 2030]{International Workshop on Software Engineering in 2030}{November 2024}{Puerto Galinàs (Brazil)}
\pdfoutput=1
\usepackage{multirow}
\usepackage{graphicx}
\usepackage{subcaption}
\usepackage{xspace}
\usepackage{algorithm}
\usepackage[noend]{algpseudocode}
\usepackage[normalem]{ulem}
\useunder{\uline}{\ul}{}
\usepackage{ragged2e}
\usepackage{array}
\usepackage{amsmath}
\usepackage{xcolor}
\usepackage{booktabs}
\usepackage[flushleft]{threeparttable}
\usepackage{enumitem}
\usepackage{pifont}
\usepackage{todonotes}
\algrenewcommand\textproc{\texttt}

\input{macros}

\AtBeginDocument{%
  \providecommand\BibTeX{{%
    \normalfont B\kern-0.5em{\scshape i\kern-0.25em b}\kern-0.8em\TeX}}}

\renewcommand\footnotetextcopyrightpermission[1]{}

\begin{document}

\title{Foundation Model Engineering: Engineering Foundation Models Just as Engineering Software}

\author{Dezhi Ran}
\affiliation{School of Computer Science, Peking University \country{China}}
\email{dezhiran@pku.edu.cn}

\author{Mengzhou Wu}
\affiliation{School of EECS, Peking University \country{China}}
\email{wmz@stu.pku.edu.cn}

\author{Wei Yang}
\affiliation{University of Texas at Dallas \country{United States}
}
\email{wei.yang@utdallas.edu}

\author{Tao Xie}\authornote{Tao Xie is with the Key Laboratory of High Confidence Software Technologies (Peking University), Ministry of Education, China, and is the corresponding author.}
\affiliation{School of Computer Science, Peking University \country{China}}
\email{taoxie@pku.edu.cn}

\thispagestyle{plain} 
\pagestyle{plain} 

\begin{abstract}
By treating data and models as the source code, Foundation Models (\textit{FMs}) become a new type of software.
Mirroring the concept of software crisis, the increasing complexity of FMs making FM crisis a tangible concern in the coming decade, appealing for new theories and methodologies from the field of software engineering.
In this paper, we outline our vision of introducing Foundation Model (FM) engineering, a strategic response to the anticipated FM crisis with principled engineering methodologies.
FM engineering aims to mitigate potential issues in FM development and application through the introduction of declarative, automated, and unified programming interfaces for both data and model management, reducing the complexities involved in working with FMs by providing a more structured and intuitive process for developers. 
Through the establishment of FM engineering, we aim to provide a robust, automated, and extensible framework that addresses the imminent challenges, and discovering new research opportunities for the software engineering field.

\end{abstract}

\maketitle

\input{overview}
\input{future}

\input{current}
\input{opportunity}
\input{conclusion}
\section*{Acknowledgments}
Tao Xie is also affiliated with the Key Laboratory of High Confidence Software Technologies (Peking University), Ministry of Education China. This work was partially supported by National Natural Science Foundation of China under Grant No. 62161146003, NSF
grant CCF-2146443, and the Tencent Foundation/XPLORER PRIZE.
Dezhi Ran was supported by National Natural Science Foundation of China under Grant No. 623B2006.

\bibliographystyle{ACM-Reference-Format}
\bibliography{ref}

\end{document}

%% file: macros.tex
\newcommand\toller[1]{\textsc{Toller}}
\renewcommand{\Comment}[1]{}

\newcommand{\llme}[0]{FME}

\newcolumntype{C}[1]{>{\centering\arraybackslash}p{#1}}

%% file: overview.tex
\section{Overview, Motivation, and Aims}\label{sec::overview}

\begin{figure}[t]
    \centering
    \includegraphics[width=\linewidth]{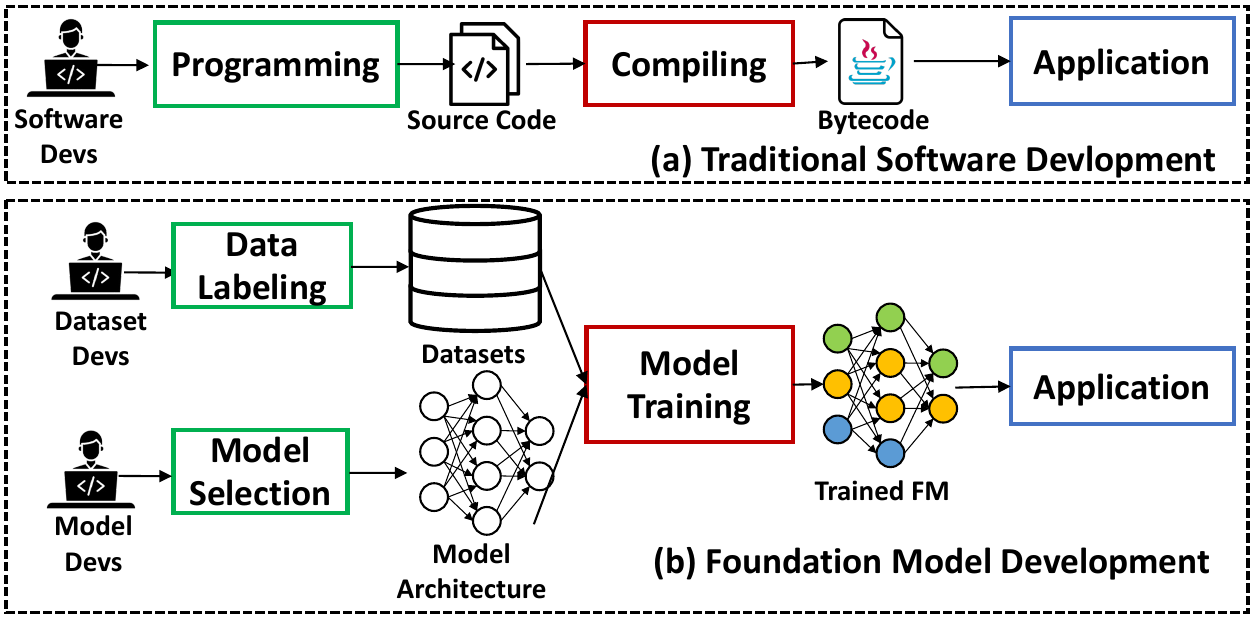}  
    \caption{An analogy between the development of traditional software and foundation models. In traditional software development, source code is manually written by human developers. The source code (e.g., cpp files) is compiled into an executable binary to perform specific tasks. In foundation model development, the ``source code" typically consists of two main components: 1) the dataset, which outlines the desired behavior, and 2) the neural network architecture, which provides a basic structure for the program, though many specifications (such as the weights) remain to be determined. Through the training process, the dataset is ``compiled" into the final foundation model, which is similar to the compiled binary of traditional software.}\label{fig::analogy}
\end{figure}

Foundation Models ~\cite{bommasani2021opportunities} (in short as \textit{FMs}) are becoming a new type of software.
An analogy between traditional software and FMs is depicted in Figure~\ref{fig::analogy}, highlighting the parallel roles of their core components.
In FM development, \textit{data} and \textit{models} play the critical role akin to source code in conventional software development.
Developers curate datasets, such as sets of example input-output pairs, to articulate the specifications for the desired FM. They then choose an appropriate network architecture and employ model training techniques such as backpropagation \cite{rumelhart1986learning}, to effectively ``compile" the dataset and network architecture into the targeted FM.

\begin{figure*}[t]
    \centering
    \begin{subfigure}[1]{0.49\textwidth}
        \centering
        \includegraphics[width=\textwidth]{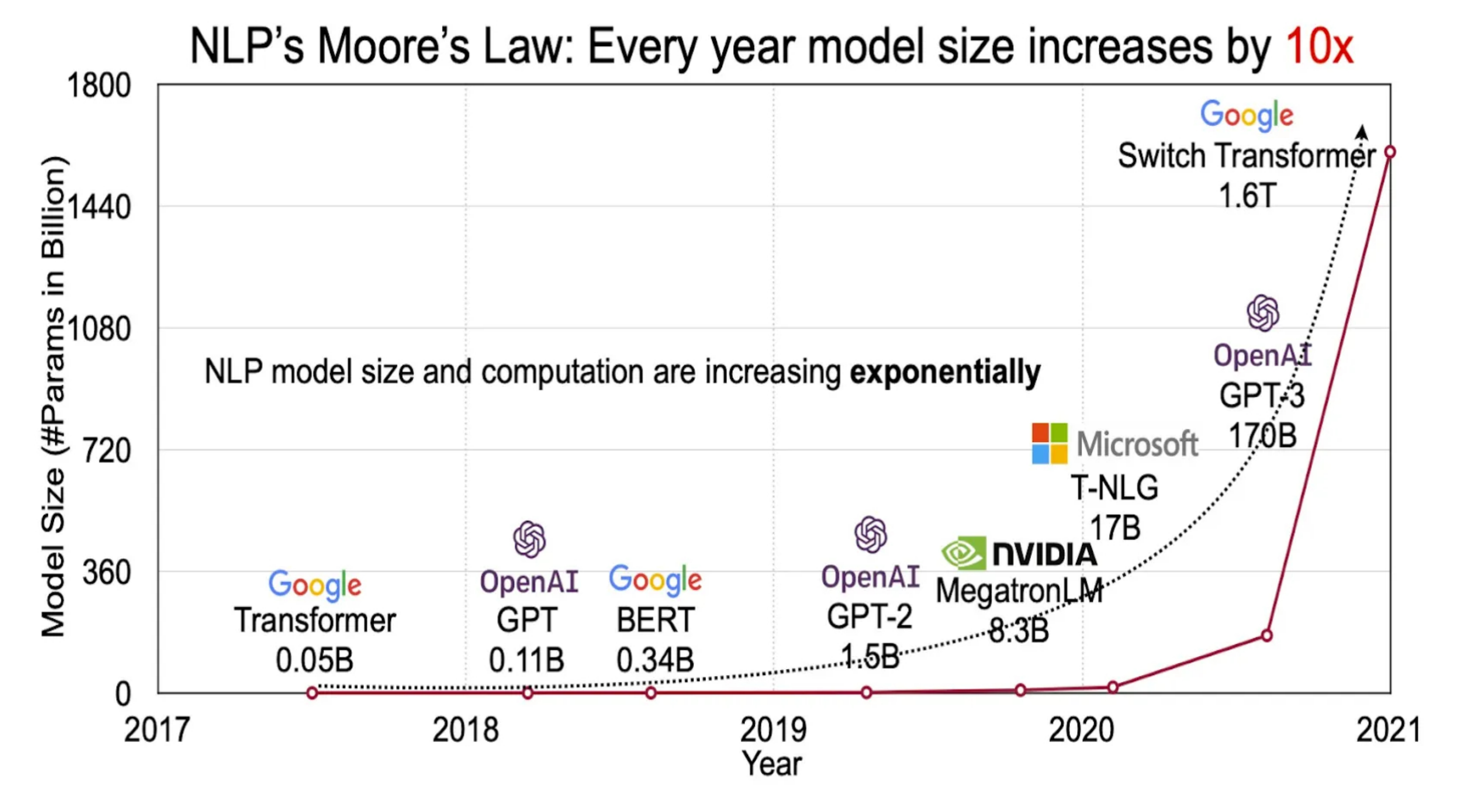}
        \caption{Increasing Complexity of FMs.}
        \label{subfig::model_complexity}
    \end{subfigure}
    \begin{subfigure}[2]{0.49\textwidth}
        \centering
        \includegraphics[width=\textwidth]{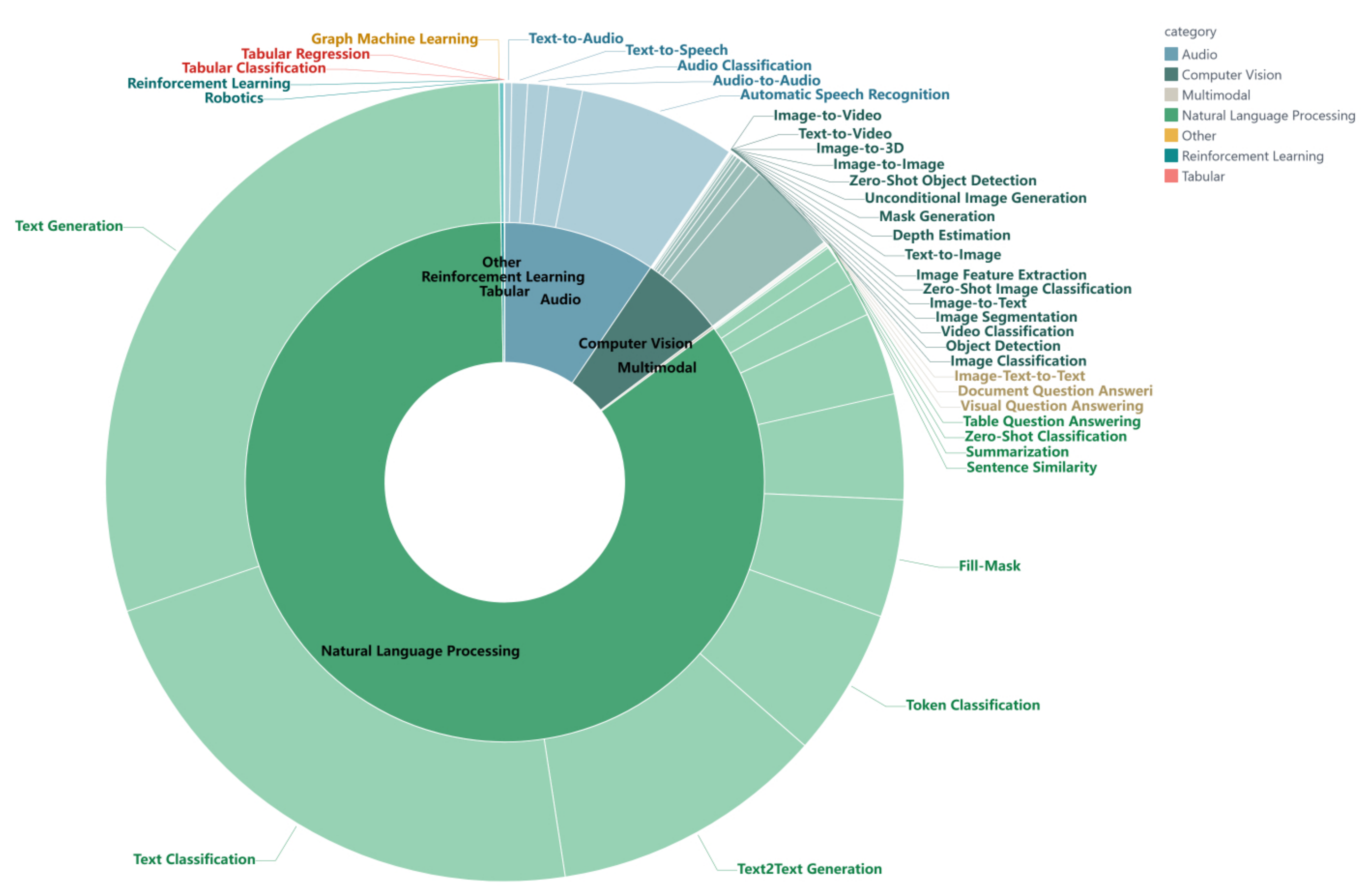}
        \caption{Incremental Development and Diverse Application Demands.}
        \label{subfig::application_diversity}
    \end{subfigure}
    \caption{
    (a) Increasing complexity of FMs (Figure~\ref{subfig::model_complexity} borrowed from ~\cite{model_size_trend}). The model size (represented by number of parameters) grows exponentially with respect to time since the invention of Transformer.
    (b) Diverse customization demand. The staggering array of models, datasets, and AI applications available on Hugging Face~\cite{hugging_face}, featuring 254,871 models with Transformer architecture, 127,462 datasets, and over 170,000 AI app demos, vividly illustrates the diverse customization demand of LLMs.
    }
    \label{fig::trend}
\end{figure*}

Given the analogy between FMs and traditional software coupled with the increasing complexity of FM development (depicted in Figure~\ref{fig::trend}), we envision that an FM crisis, mirroring the concept of the software crisis~\cite{dijkstra1972humble}, will surface as a tangible concern over the coming decade.
This potential crisis can be evidenced across four major aspects:
\begin{itemize}
\item \textbf{Increasing FM complexity.} 
Since the invention of Transformer~\cite{vaswani2017attention}, the complexity of FMs (represented by the number of parameters) has been increasing at an exponential rate.
This growing complexity raises significant challenges not only in training and managing these FMs but also in managing the datasets that underpin them. 
Specifically, as datasets grow in size to unprecedented levels, the tasks of cleaning, labeling, and managing data at such a scale become increasingly challenging.

\item\textbf{Continuous FM evolution.} 
FMs are in a state of continuous and fast evolution, driven by the integration of new data, the emergence of new requirements, and the implementation of bug fixes.
For example, the average time between updates for OpenAI's models is 2 weeks~\cite{openai_update_freq}, more frequent than that of Linux, whose mainline kernels are updated every 9-10 weeks~\cite{linux_update_freq}.
Beyond the updates to the base FM, developers often finetune these models within their specific application domains. 
When the base FM updates, particularly security-related fixes, it necessitates a corresponding update to the finetuned FMs. 
It is challenging to efficiently and cost-effectively manage the evolution of these finetuned FMs, catching up with the base model updates while maintaining their domain-specific enhancements.

\item\textbf{Diverse customization demand.} 
The era of FMs is characterized not only by their technological advancements but also by the broad spectrum of customization demands.
On Hugging Face~\cite{hugging_face}, there are 254,871 models with Transformer architecture, 127,462 datasets, and over 170,000 AI app demos across various application domains.
This diverse ecosystem of models, datasets, and applications is a clear indication that the future of LLMs lies in their ability to be customized.
During the customization, ensuring data privacy, model interpretability, and bias elimination requires innovations of data management and model management.

\item\textbf{Multi-agent collaboration.}
The development of FMs involves multi-agent collaboration among data scientists, model developers, and application domain experts. 
This collaboration introduces significant hurdles, primarily due to differing objectives, terminologies, and methodologies across disciplines.
The specialized language used by data scientists and engineers may not align with the domain-specific knowledge and practical insights of application domain experts, potentially leading to misunderstandings and delays in project timelines. 
Innovative techniques are needed for minimizing the barriers posed by interdisciplinary work and effective collaboration.
\end{itemize}

Inspired by the role that software engineering has played in addressing software crisis~\cite{dijkstra1972humble}, this paper outlines our vision of \textit{foundation model engineering} (depicted in Figure~\ref{fig::infra_overview}), a strategic response to the anticipated FM crisis.
Foundation model engineering aims to mitigate potential issues in FM development and application through the introduction of declarative, automated, and unified programming interfaces for both data and model management, reducing the complexities involved in working with FMs by providing a more structured and intuitive process for developers. 
The key components of our proposed foundation model engineering include:
\begin{itemize}
    \item \textbf{Data management with weak supervision.}
    Recognizing the importance of high-quality data in training effective FMs, we advocate for advanced data management strategies that leverage weak supervision. 
    This approach allows for the efficient labeling and curating of vast datasets by combining limited amounts of labeled data with large quantities of unlabeled data, using algorithms to infer labels and improve data quality. This method significantly reduces the time and resources required for data preparation.
    \item \textbf{Model management with workflows and continuous integration.}
    Effective model management is essential for the scalable and sustainable development of FMs. We envision the implementation of workflows that encompass model development, training, evaluation, and deployment processes, automating routine tasks and ensuring optimal practices.
    We also envision a distributed version control system like Git~\cite{loeliger2012version} to track the update of FMs, manage model branches, and resolve conflicts of model updates.

    \item \textbf{Programmatic FM development with declarative specifications.} To further simplify the engineering of FMs, we envision unified and declarative APIs that abstract away the underlying complexities of model and data management. These APIs allow developers to specify what they want to achieve in a high-level language, without needing to provide detailed instructions on how to accomplish these tasks. This not only makes the process more accessible to a wider range of users, including those with less technical expertise but also accelerates the development cycle by enabling quicker iterations and refinements of models.

\end{itemize}

Through the establishment of foundation model engineering as envisioned in this paper, we aim to provide a robust, automated, and extensible framework that addresses the imminent challenges and opportunities presented by the rapid advancement of FMs. 
By equipping developers with the tools and methodologies to effectively and efficiently manage data and models, we can enhance productivity and responsible use of these powerful FMs, and discover research opportunities for the software engineering community in the FM era.
In summary, our aims are to:
\begin{itemize}
    \item envision Foundation Model (FM) engineering for resolving the potential ``FM crisis" in the next decade.
    \item architect FM engineering in terms of data management, model management, and declarative programming interfaces. 
    \item research on tools, techniques, and methodologies for improving and automating FM engineering.
\end{itemize}

%% file: future.tex
\section{Envision of FM Engineering}\label{sec::future}

\begin{figure}[t]
    \centering
    \includegraphics[width=\linewidth]{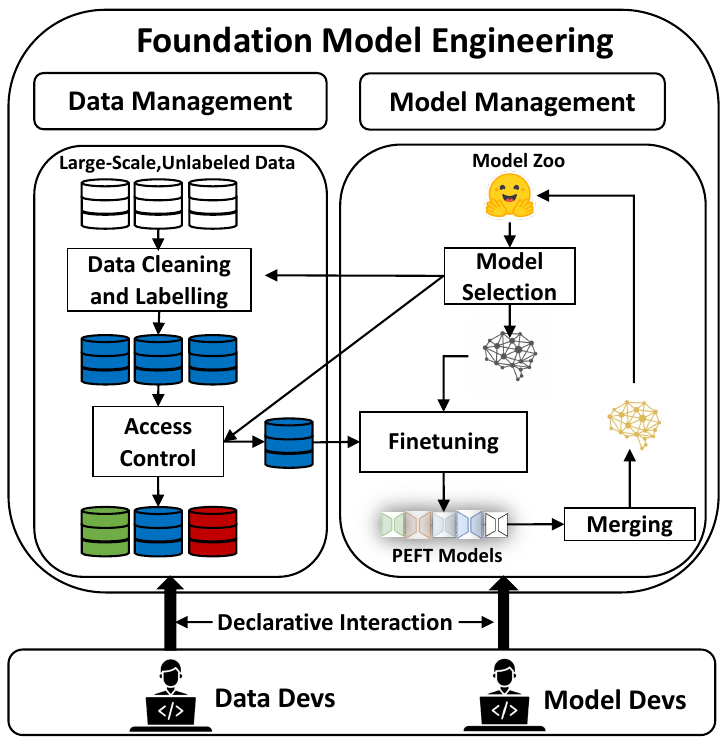}  
    \caption{Overview of Foundation Model (FM) Engineering.}
    \label{fig::infra_overview}
\end{figure}

\subsection{Overview of Foundation Model Engineering}
Figure~\ref{fig::infra_overview} presents the overview of the envisioned Foundation Model Engineering (in short as \textit{\llme{}}), which manages resources, abstracts common operations, and provides APIs for related developers to engineer data and model in declarative ways.

\noindent\textbf{Data management.}
In the landscape of \llme{}, data management emerges as a key element. The ease with which machines generate massive volumes of data presents a unique challenge, especially in ensuring full coverage across enormous public or private information sources where collection complexities multiply. The criticality of data relevance cannot be overstated; amidst the deluge, distinguishing valuable data for analysis and decision-making becomes crucial. Moreover, the effort to process and refine this raw data into a form that is both accessible and actionable is a formidable task that \llme{} tackles head-on.

At the core of \llme{}'s strategy is the fundamental principle that acquiring the appropriate data is critical to the success of the entire operation. It is widely recognized that simple models built on well-curated datasets often surpass their complex counterparts that are fed skewed or incomplete data. An integral part of this approach includes rigorous auditing and inspection of data pipelines, which safeguards data integrity and ensures that processing aligns with predefined goals and compliance requirements.

Once raw data is collected, the next phase involves transforming this data into meaningful signals within \llme{}. This transformation is a multi-step procedure that commences with data wrangling for cleaning and structuring, followed by aggregation to distill summary insights. Anomaly detection algorithms sift through to highlight irregularities, while pattern-matching and linear regression inform on current trends and future directions. The process reaches its end with the deployment of advanced machine learning models that extract complex patterns and forecasts, thereby deepening the comprehension of the data's story.

\llme{} maintains the data assets for model training and finetuning.
Data developers can declare data labelling and cleaning functions with high-level intention descriptions and weak-supervision, such as specifying exemplar data-label pairs.
\llme{} takes the schema as input, selects an appropriate model for parsing and understanding the schema (typically a code generation model or SQL generation model), and generates data labeling or cleaning functions to perform the data labelling or cleaning at scale.
In addition to data labeling, \llme{} also implements access control to data assets to assure the data access conforms to legal policies as well as user intentions.

\noindent\textbf{Model management.} 
\llme{} champions a novel paradigm for machine learning, paralleling the collaborative dynamism of open-source software development. This initiative seeks to transform the lifecycle of foundation models from static entities to evolving constructs, continuously refined through community contributions. Unlike traditional open-source software, which thrives on collective inputs and evolution, foundation models often see their development halt post-release. To bridge this gap, \llme{} is cultivating a culture where machine learning models are not just released but are actively developed, enhanced, and adapted through collaborative efforts.

Central to this culture shift is the strategic facilitation of efficient change communication and contribution integration, steering clear of the impracticalities of transferring voluminous parameters characteristic of contemporary models. Leveraging insights like Fisher information~\cite{ly2017tutorial}, \llme{} focuses on pinpointing and updating specific subsets of parameters. This targeted approach enables substantial performance enhancements without the heft of large-scale data transfers.

In the spirit of collaborative software engineering, the exploration of merging contributions from disparate sources stands as a testament to the power of collective intelligence. By integrating the diverse expertise of independently crafted models, \llme{} paves the way for new capabilities and amplified performance in specialized tasks.

The push towards modularity in machine learning is informed by the ``mixture of experts" architectural paradigm~\cite{muqeeth2023soft, zhang2022skillnet, du2022glam, zadouri2023pushing}, where specialized sub-networks synergize through adaptive routing, facilitating the backpropagation of discrete choices. Such an arrangement empowers models to assimilate domain-specific knowledge with unprecedented efficiency.




Lastly, to facilitate the version control and integration of updates, the concept of git for models, is being developed~\cite{kandpal2023git}. This system aims to track parameter updates, support efficient merging, and integrate seamlessly with existing workflows, marking a significant step towards a more collaborative machine learning ecosystem.


\llme{} also automates the model finetuning process, where developers specify tasks and declare accessible data assets. The system intelligently selects suitable models, identifies finetuning data from the available pool, and executes finetuning with auto-adjusted hyperparameters. Upon completion,
 \llme{} merges the redundant models, resolves conflicts during merging, and returns the merged model for subsequent usage.



\subsection{Data Management}

\begin{figure}[t]
    \centering
    \includegraphics[width=\linewidth]{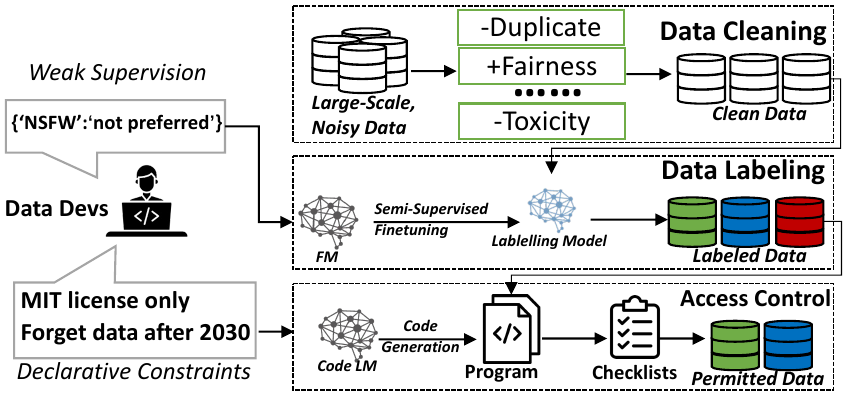}  
    \caption{Overview of Data Management.}
    \label{fig::data_management_detai}
\end{figure}
As shown in Figure~\ref{fig::data_management_detai}, we envision an integrated environment and declarative interfaces that support data management with minimal human efforts.  
Data is the user interface to ``program" FMs, and a user-friendly interface is expected to fulfill four major requirements.
First, FMs are trained on vast datasets, comprising billions of words and documents from the Internet, books, articles, and more. 
The quality, diversity, and size of the training data directly affect the LLM's ability to understand and perform tasks ranging from dialog system to code generation.
Second, labelled data are required to adapt FMs to specific downstream tasks or non-functional properties such as human preference alignment~\cite{ouyang2022training}.
Third, data sourcing, access control and unlearning are crucial for data management to conform to legal policies and user privacy requirements.
Finally, the goal is to fulfill the above requirements in an automated streamline with minimal human efforts.

To fulfill the preceding requirements, we envision that the data management module of \llme{} consist of the following four modules.

\noindent\textbf{Data cleaning.}
The data cleaning module, critical in ensuring data integrity and quality for FMs, embraces a high-level, declarative framework. Within this framework, users specify the desired characteristics of clean data, guiding the module to automatically pinpoint and rectify inaccuracies, inconsistencies, and redundancies in the datasets according to these specifications. 
The module is enriched with an interactive feedback loop. This crucial feature allows users to review and validate the automated cleaning actions undertaken by the system, facilitating an iterative refinement process. Through this dynamic interaction, users can fine-tune the cleaning criteria, ensuring it resonates with the unique aspects of the data and their specific needs. 
By employing automated tools and algorithms, this module can detect anomalies, filter out irrelevant or sensitive information, and standardize data formats. This step is crucial for reducing noise in the training data, thereby enhancing the FM's learning efficiency and effectiveness in understanding complex language patterns and generating coherent outputs.


\noindent\textbf{Data labelling.}
Data labelling focuses on annotating the training data with informative tags or labels that define the context or the desired outcome of the data. This is particularly important for supervised learning tasks where the FM needs to recognize patterns or generate responses based on specific inputs. Leveraging automated data labeling strategies with human specifications, establish high-level rules or heuristics, facilitating the automatic creation of labels over large datasets.
This method is strengthened by incorporating specialized knowledge through labeling functions and using weak supervision techniques to utilize a broad range of information sources for deducing labels.
In addition, the module iteratively enhances label accuracy by leveraging model-driven insights to rectify initial ambiguities or errors introduced by heuristic rules or noisy labels. 
This module enables the customization of FMs for specialized applications, from sentiment analysis to personalized content creation, by aligning the model's outputs with human preferences and task-specific requirements.

\noindent\textbf{Access control.}
Data sourcing must comply with copyright laws, privacy regulations, and ethical standards. The use of publicly available data, proprietary datasets, and user-generated content requires careful examination.
Access control mechanisms are implemented to manage who can view or use the data, ensuring that only authorized users or systems have the ability to access or modify the datasets. 
By incorporating robust authentication, authorization, and auditing processes, the access control module safeguards sensitive information and prevents unauthorized data breaches, thereby fostering trust in the LLM's development and deployment processes.

\noindent\textbf{Weak supervision.}
The weak supervision module aims to reduce the reliance on extensively labeled datasets by utilizing less precise labels that can be generated more easily or derived from heuristic rules and external knowledge sources. 
This approach allows for the rapid scaling of training data while managing resource constraints and minimizing manual labelling efforts. 
Through advanced algorithms and models that can learn from weakly supervised data, this module supports the efficient training of FMs across a broader range of tasks and domains, accelerating the model's adaptability and performance improvement.

\subsection{Model Management}

\begin{figure*}[t]
    \centering
    \includegraphics[width=0.9\textwidth]{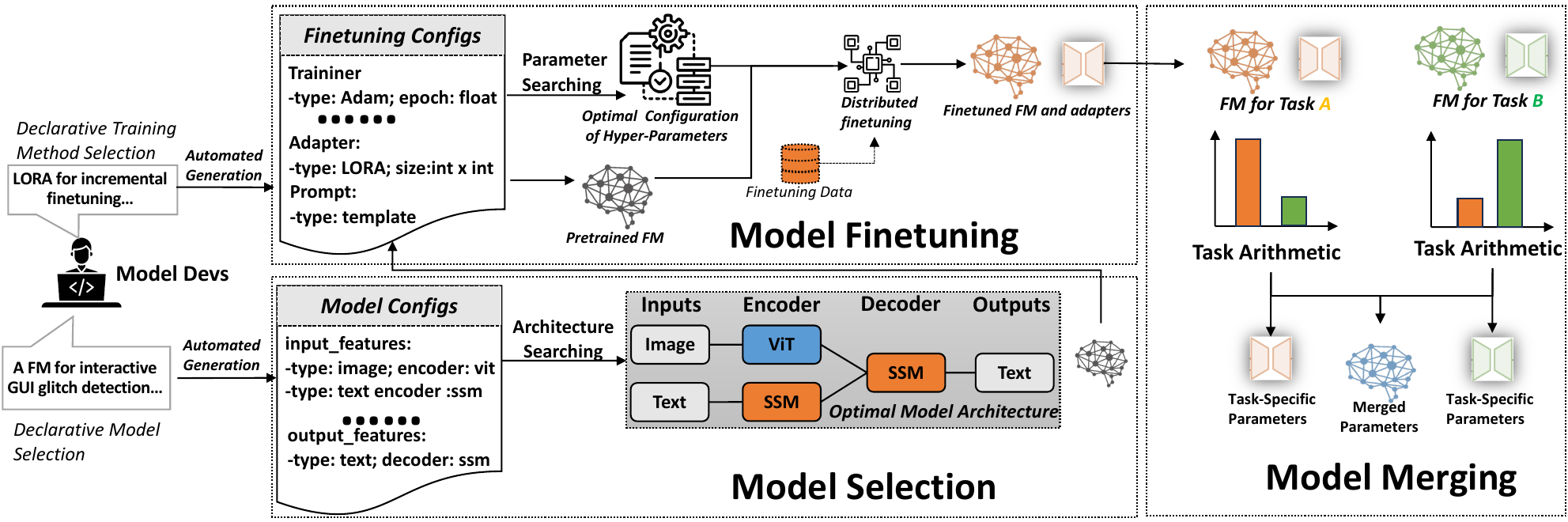}  
    \caption{Overview of Model Management.}
    \label{fig::model_management_detail}
\end{figure*}

As shown in Figure~\ref{fig::model_management_detail}, we develop an integrated environment and distributed version control system that supports the development, evolution and deployment of FMs. 

Nowadays, model updates rarely start from retraining from scratch but instead involve incremental fine-tuning where only a small fraction of parameters are changed.
Considering that different applications may finetune their own FMs from the same pretrianed FM, maintaining separate FMs for different application encounters similar problems as code cloning problems common in traditional software engineering, reducing software reliability and maintainability.
Given that different users have different fine-tuning data, leading to different or even conflict parameter updates, maintaining these FMs can be challenging for developers to manually check and update the FMs.

\begin{figure*}[t]
    \centering
\includegraphics[width=0.85\textwidth]{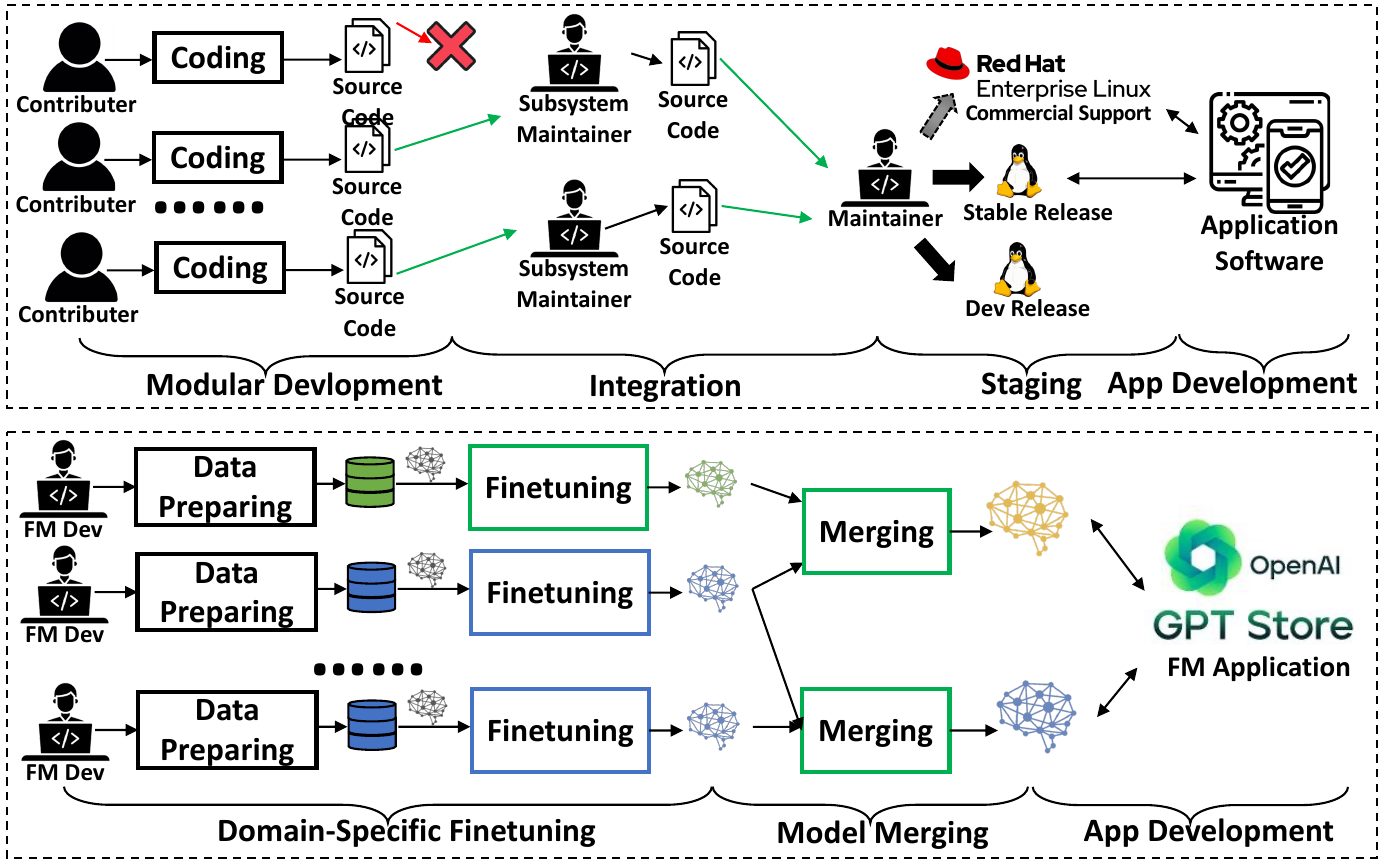}  
    \caption{Envisioned FM Update Process with an Analogy with Linux Update Process.}
    \label{fig::model}
\end{figure*}

Inspired by Git~\cite{loeliger2012version}, we envision a distributed version control system of foundation models (shown in Figure~\ref{fig::model}), providing platform support for the community-driven development, evolution, and continuous integration of FMs, The system offers a unified software development and version management abstraction for both new applications and the supporting platform.

\noindent\textbf{Model selection.}
The model selection module empowers users to identify the most suitable FMs for their specific applications based on performance benchmarks, compatibility, and previous usage outcomes. 
For example, the Composition of Experts (CoE)~\cite{coe} is used in model selection to aggregate multiple specialized models to improve overall performance and accuracy. For example, in a CoE system, there could be distinct models expert in language understanding, image recognition, and sentiment analysis. When faced with a complex task that involves understanding text within images and gauging sentiment, the CoE framework would select and combine the outputs of these expert models. This modular approach allows for targeted fine-tuning of each expert model, ensuring that the collective output is both accurate and efficient in handling the task at hand.

\noindent\textbf{Model finetuning.}
The model finetuning module offers a flexible and user-friendly toolkit for customizing FMs to meet the unique demands of diverse applications. It simplifies the process of applying incremental updates, adjusting parameters, and integrating new data, thereby enabling personalized model optimization without the need for extensive machine learning expertise. The finetuning module provides an intuitive toolkit, enabling users to tailor Foundation Models for diverse needs with ease. It supports collaborative, large-scale, distributed learning environments, where contributors work on separate data without sharing, ensuring data privacy and ownership while optimizing model performance through a central minimal-computation repository. This distributed setup allows for parallel training tasks on shared servers, maintaining parameter ownership and preventing task interference, streamlining the path to personalized model optimization.

\noindent\textbf{Model merging.}
The process of model merging after finetuning involves a collaborative and incremental approach, similar to practices in open-source software development.
Addressing the challenges of divergent fine-tuning efforts, the model merging module incorporates sophisticated algorithms to harmonize changes from multiple sources.
This method allows for efficient communication of updates between contributors and a central repository, focusing on updating only a selected subset of parameters based on their significance, as determined by measures like Fischer information~\cite{ly2017tutorial, sung2021training}. This approach ensures that updates are manageable in size, reducing communication costs and complexity. The ultimate goal is to combine the strengths of independently trained models, preserving the benefits of each, to enhance the overall performance and capabilities of the merged model in a distributed and collaborative learning environment.
This ensures consistency, mitigates conflicts, and maintains the integrity of FMs across different applications, significantly reducing the maintenance burden and promoting collaborative improvements.

\noindent\textbf{Model deployment.}
The model deployment module streamlines the process of rolling out updated or newly finetuned FMs into production environments. The deployment process begins with user requests, which are systematically queued. These requests are then managed by a deployment setup, often running on a Kubernetes environment, where the FM is loaded into memory within containers organized into pods. Within this deployment, there are two primary types of model weights: the base model weights, which form the core parameters of the FM, and adapter weights, which are a smaller set of parameters that allow for model fine-tuning.

One of the central challenges in deploying fine-tuned FMs is the significant resource requirements, particularly in terms of GPU usage. Each new user or task traditionally necessitates a new pod and GPU, leading to potentially excessive resource consumption. However, a crucial insight is that most of the model weights across different deployments remain identical, with variations primarily in the adapter weights. This realization opens up possibilities for sharing the base model across multiple adapters, thereby optimizing resource use.

Addressing this challenge, we envision a system with several innovative components. Firstly, it employs dynamic loading of adapter weights, allowing the system to serve multiple user requests by only loading the necessary adapter weights alongside the base model weights into memory. This approach significantly reduces the need for additional resources per user or task. Secondly, the system incorporates a multi-tier weight cache to manage adapter weights efficiently. This cache includes a GPU cache for actively used adapters, a CPU cache for adapters awaiting activation, and an idle tier for adapters not currently in use, with their weights stored on ephemeral disk storage for potential future requests.

Another key innovation that we can adopt here is continuous multi-adapter batching, an extension of the continuous batching concept~\cite{lorax}. This technique allows the system to process requests from multiple adapters together in the same batch, significantly improving throughput and efficiency. The batching algorithm central to this system prioritizes adapters based on request timestamps and employs a cycle time parameter to manage the swapping of adapters in and out of the active set, striking a balance between throughput and latency.
The model deployment module ensures smooth transition, minimizes downtime, and facilitates continuous delivery, allowing developers and users to leverage the latest advancements with ease and confidence.

Together, these components form a comprehensive ecosystem that not only simplifies the management of FMs but also accelerates the pace of innovation, fostering a collaborative and dynamic environment for the advancement of intelligent applications.

%% file: current.tex
\section{State of the Practice}\label{sec::current}

\subsection{Data-Centric Machine Learning}
\citet{kaplan2020scaling} reveals that improving in model architectures usually offer limited benefits. 
In contrast, the efficacy of data utilization is becoming the cornerstone of advancing model performance given the increasing dataset scale enabled by self supervised learning techniques~\cite{devlin2018bert, radford2019language}.

\noindent\textbf{Data cleaning.}
Data cleaning is the process of addressing errors, duplications, and incompleteness in datasets by modifying, adding, and deleting data. 
Holoclean~\cite{rekatsinas2017holoclean} utilizes a variety of methods including heuristic rules (such as integrity constraints), external knowledge, and quantitative statistics to integrate multiple data sources into a probabilistic model, identifying and correcting errors in datasets. Picket~\cite{liu2022picket} employs self-supervised deep learning models to identify and remove corrupted data without the need for human supervision. 
\citet{neutatz2021cleaning} found that the benefits of data cleaning largely depend on the application, leading to the proposal of end-to-end, application-driven, holistic data cleaning approaches.
ActiveClean\cite{krishnan2016activeclean} combines data cleaning with active learning, prioritizing the cleaning of data that could potentially impact model performance in specific application domains. 

\noindent\textbf{Data programming with weak supervision.}
The need for large labeled datasets to optimally train modern machine learning models presents a major bottleneck, due to the high costs and time needed for expert manual annotation.
In response, active learning strategies\cite{settles2009active} streamline this process by selectively engaging experts to label data of maximal utility—such as instances at the fringes of classification models—thereby amplifying model efficacy with diminished input. Parallelly, semi-supervised learning\cite{van2020survey} capitalizes on a modest quantum of labeled data supplemented by unlabeled data to enhance model accuracy, effectively economizing on the need for extensive labeled datasets. Weak supervision approaches harness cost-effective techniques for gathering lower-quality labeled data through avenues like Crowdsourcing\cite{karger2011iterative}, Distant Supervision\cite{mintz2009distant}, or heuristic rules, markedly alleviating the reliance on manual labeling. The Snorkel system\cite{ratner2017snorkel} enables users to employ labeling functions that encapsulate heuristic methods, combines different weak supervision sources to generate a probabilistic distribution of labels.

\subsection{Incremental Model Training}

\noindent\textbf{Parameter-Efficient Fine-Tuning (PEFT).} The extensive parameter set of FMs renders their fine-tuning both computationally expensive and storage-intensive. PEFT techniques offer a solution by fine-tuning of a fraction of their parameters,
The compact nature of specialized PEFT modules facilitates their dissemination within the community, as evidenced by the availability of over 20,000 adapters on the Hugging Face Model Hub, all of which are based on the PEFT framework~\cite{peft}.

PEFT strategies are predominantly categorized into three distinct approaches. 
First, \textit{additive methods} entail the integration of additional parameters into the original Large Language Model (LLM) architecture, with the fine-tuning process focusing exclusively on these new parameters. Notably, \citet{houlsby2019parameter} introduced fully-connected networks as adapter modules within the transformer architecture, after the attention and Feed-Forward Network (FFN) layers. Prompt tuning\cite{lester2021power} and prefix tuning\cite{li2021prefix} incorporate task-specific vector sequences at the input layer and throughout various layers of the LLM, respectively, with fine-tuning achieved through the adjustment of these vector parameters. 
Second, \textit{selective methods} involves the selective training of a subset of the LLM's parameters. BitFit\cite{zaken2021bitfit} fine-tune only the bias parameters, while DiffPruning\cite{guo2020parameter} employs an L0-norm to train a sparse weight matrix. Freeze and Reconfigure\cite{vucetic2022efficient} and FishMask\cite{sung2021training} identify and train crucial model parameters based on L1-distance and Fisher information, respectively.
Third, \textit{reparametrization-based methods} modifies the original model's parameter matrix into a more tractable low-rank format for training purposes. Intrinsic SAID\cite{aghajanyan2020intrinsic} utilizes the FastFood transform for reparameterizing model weight updates. LoRa\cite{hu2021lora} decompose the weight matrix updates into products of low-rank matrices, with KronA\cite{edalati2022krona} employing Kronecker product for matrix factorization. AdaLoRa\cite{zhang2022adaptive} adopts Singular Value Decomposition (SVD) for parameter matrix decomposition, prioritizing resources based on the significance of different weight matrices. Additionally, innovative integrations such as SparseAdapter\cite{he2022sparseadapter}, MAM Adapters\cite{he2021towards}, UniPELT\cite{mao2021unipelt}, Compacter\cite{karimi2021compacter}, and S4\cite{chen2023parameter} amalgamate various PEFT methodologies to enhance efficiency and adaptability.

\noindent\textbf{Modular training for multi-task learning.}
\citet{ilharco2022editing} introduces the concept of a task vector to delineate the shifts within the model's parameter space consequent to fine-tuning for a specific task. He further elucidates that performing arithmetic operations on this task vector facilitates the processes of forgetting, constructing multi-task models, and task analogies. In the quest to refine model merging capabilities leveraging the task vector, \citet{matena2022merging} applies Fisher information weighting, \citet{jin2022dataless} frames the challenge as an optimization quandary and resolves it via linear regression, while \citet{yadav2024ties} mitigates interference by eliminating superfluous parameters and reconciling symbol discrepancies. 
\citet{choshen2022fusing} and \citet{don2022cold} engage in iterative fine-tuning of the Fundamental Model (FM) across diverse tasks, subsequently averaging the weights to enhance the FM. Further, an amalgamation of PEFT techniques with model merging strategies is explored\cite{lv2023parameter, huang2023lorahub, zhang2024composing, chronopoulou2023adaptersoup, ponti2023combining, pfeiffer2020adapterfusion}. 
\citet{ponti2023combining} posits that each task is linked to a spectrum of skills, with each skill mirrored by an adapter, and devises a routing function to allocate skills per task. AdapterSoup\cite{chronopoulou2023adaptersoup} employs task textual similarity and clustering techniques to identify auxiliary tasks conducive to the target task, facilitating the merger of pertinent adapters. LoRaHub\cite{huang2023lorahub} adopts a gradient-free optimization strategy to fine-tune LoRa model merging, guided by few-shot examples of the target task.

A noteworthy trend in recent research\cite{muqeeth2023soft, zhang2022skillnet, du2022glam, zadouri2023pushing} is the development of mixture-of-experts models to tackle the multi-task conundrum. Within such models, a selective activation of a subset of experts is triggered at each layer contingent on the input, thereby focusing inference and training efforts solely on the activated experts. This approach has been shown to yield superior performance, particularly when models are extended to tasks beyond their initial training scope.

%% file: opportunity.tex
\section{Research Opportunities}\label{sec::opportunity}
To fulfill our vision of FM engineering, numerous research opportunities arise, which can be categorized into three-fold.

\subsection{Declarative FM Engineering}
Declarative FM engineering represents a paradigm shift in the development of Large Language Models, emphasizing the specification of what the model should achieve rather than how it achieves it. This approach, rooted in the principles of declarative programming, offers a more intuitive and efficient methodology for designing, training, and deploying FMs. It invites a wealth of research opportunities aimed at simplifying the complex process of FM engineering, making it more accessible and adaptable to a broader range of applications and developers.

\noindent\textbf{High-level model specification languages.} Developing high-level, domain-specific languages for FM engineering that allow developers to specify the desired outcomes, constraints, and behaviors of the model in an abstract manner. Research in this area could focus on creating intuitive syntax and semantics that encapsulate the complexities of neural network architectures, training procedures, and data processing pipelines.

\noindent\textbf{Automated model synthesis.} Building on high-level specifications, automated model synthesis involves research into algorithms and systems capable of translating these abstract descriptions into concrete, optimized FM architectures. This includes selecting appropriate neural network components, configuring layers and connections, and determining optimal training strategies based on the specified objectives and constraints.

\noindent\textbf{Constraint-based optimization}. Investigating methods for incorporating various types of constraints (e.g., performance, fairness, privacy) directly into the FM training process in a declarative manner. This research area would explore optimization techniques that can balance multiple objectives and adhere to specified constraints, ensuring that the resulting models align with ethical guidelines and application-specific requirements.

\noindent\textbf{Verification and validation.} Given the abstract nature of declarative specifications, developing robust verification and validation methods is crucial to ensure that the synthesized FMs faithfully represent the intended outcomes and adhere to all specified constraints. This includes formal verification techniques to prove properties about the models and empirical validation approaches to evaluate their performance and behavior in real-world scenarios.

\subsection{Fine-grained Data Management}

\noindent\textbf{Granular permissions.} Developing systems that enable granular control over who can access specific datasets or parts of datasets is crucial. This includes defining roles and permissions at a detailed level, allowing for precise management of data access based on the user's role, the nature of the project, and the sensitivity of the data. Research into models that can dynamically adjust permissions in response to changing project needs or data sensitivity levels could significantly enhance data security and governance.

\noindent\textbf{Decentralized access control.} With the rise of decentralized technologies, investigating decentralized access control models, such as those based on blockchain, could offer new ways to manage data assets securely and transparently. These models could provide immutable, verifiable logs of data access and changes, enhancing trust and compliance.

\noindent\textbf{Automated compliance checks.} Given the complex web of data protection laws globally, developing automated systems for compliance checks during data access can help organizations navigate legal requirements more efficiently. Research into AI-driven compliance advisors that can interpret and apply legal rules in real-time during data access decisions could greatly reduce the burden of legal compliance.

\subsection{Automating Model Management}

The need to evolve pre-trained models parallels the dynamic nature of open-source software development, emphasizing  the critical requirement for robust infrastructure support for incremental, collaborative, and agile enhancement of foundation models. Pre-trained models, much like any sophisticated software, may require updates for a variety of critical reasons: enhancing their performance through extended training or alternative data, rectifying problematic outputs such as noise or offensive content, and addressing privacy concerns related to memorized data.

However, the current practice largely involves models remaining static post-release, awaiting replacement by a completely new version, rather than undergoing continuous refinement. This approach  contrasts with the evolutionary trajectory of open-source projects, such as popular programming languages such as Python or Java. Had these languages remained unchanged since its initial release, it would lack many now-essential features and fixes, all contributed by a diverse community of developers. These enhancements, often seamlessly integrable into existing codebases, are facilitated by a well-established ecosystem of development tools and practices, including version control, continuous integration, and package management.

Adopting a similar framework for the development of foundation models would not only enable their continual improvement but also ensure their relevance and utility in an ever-evolving technological landscape. 
This necessitates the establishment of standardized protocols for model updates, a transparent versioning system to manage changes, and comprehensive guidelines for adapting existing applications to updated models. 

The success of ecosystems such as the one found in the Linux operating system, a predominant force within the open-source community, is largely attributed to an extensive array of both built-in and externally contributed libraries. The effective management of these libraries, including their installation, removal, and dependency resolution, alongside version control for tracking developmental progress, is crucial.
Within the Linux ecosystem, `Git`, a distributed version control system, has been instrumental in fostering collaborative and parallel development, thereby accelerating the growth and evolution of the Linux ecosystem.

The use of ``Git" predominantly focuses on line-level changes during code comparison, particularly in processes like merging and rebasing, without delving into the semantic meanings of the text. 
However, with the advent of agent development using natural language, as highlighted in earlier discussions, adapting traditional version control systems to handle natural language brings forth unique challenges. Natural languages, despite being governed by grammatical rules, often exhibit a loosely coupled relationship among the diverse expressions utilized by different individuals. This aspect can lead to natural language statements that, while semantically equivalent, may vary significantly in their phrasing.

%% file: conclusion.tex
\section{Conclusion}
In conclusion, this paper envisions foundation model engineering, aiming to streamline the development of foundation models through innovative infrastructure software and methodologies. 
By simplifying data and model management and emphasizing automated, declarative interfaces, we envision a future where LLMs are more accessible, efficient, and ethically developed. This approach not only promises to accelerate innovation within the machine learning field but also ensures that the profound benefits of LLMs can be leveraged across various sectors, contributing positively to societal advancement. As we advance, collaborative and conscientious efforts will be key to realizing the full potential of these technologies in a responsible and beneficial manner. 